# Classification of Solar Radio Spectrum Based on Swin Transformer

Jian Chen [1], Guowu Yuan [1,2,*], Hao Zhou [1], Chengming Tan [3,4] and Lei Yang [1] and Siqi Li [1]

[1] School of Information Science & Engineering, Yunnan University, Kunming 650500, Yunnan, China; jianchen@mail.ynu.edu.cn (J.C.); zhouhao@ynu.edu.cn (H.Z.); yanglei@mail.ynu.edu.cn (L.Y.); sqli@mail.ynu.edu.cn (S.L.)
[2] CAS Key Laboratory of Solar Activity, National Astronomical Observatories, Beijing 100012, China
[3] State Key Laboratory of Space Weather, National Space Science Center, Beijing 100864, China; tanchengming@nssc.ac.cn
[4] School of Astronomy and Space Sciences, University of Chinese Academy of Sciences, Beijing 100049, China
* Correspondence: gwyuan@ynu.edu.cn

**Abstract:** Solar radio observation is a method used to study the Sun. It is very important for space weather early warning and solar physics research to automatically classify solar radio spectrums in real time and judge whether there is a solar radio burst. As the number of solar radio burst spectrums is small and uneven, this paper proposes a classification method for solar radio spectrums based on the Swin transformer. First, the method transfers the parameters of the pretrained model to the Swin transformer model. Then, the hidden layer weights of the Swin transformer are frozen, and the fully connected layer of the Swin transformer is trained on the target dataset. Finally, parameter tuning is performed. The experimental results show that the method can achieve a true positive rate of 100%, which is more accurate than previous methods. Moreover, the number of our model parameters is only 20 million, which is 80% lower than that of the traditional VGG16 convolutional neural network with more than 130 million parameters.

## 1. Introduction

The Sun is the closest star to the Earth, and the light and heat it provides are the source of human survival and activities on the Earth. The Sun's violent activity and space weather changes affect human survival. Solar bursts are a sporadic component of solar radio emission connected with the flare energy release. A solar radio spectrometer can observe the solar radio emission intensity in the radio band and is the main equipment for studying a solar burst. If the solar radio spectrum can be processed automatically and solar burst events can be classified in real time, it is of great value to solar physics research and space weather early warning.

In the late 1960s and early 1970s, John Paul Wild's team built and operated the world's first solar radio spectrometer and the subsequent Culgoora radio heliograph. Since then, an increasing number of solar radio spectrometers have been built. In recent years, many researchers have studied the automatic processing of the observed solar radio spectrum. Ruizhen Zhao analyzed the features of the NeighShrink thresholding function, and a new wavelet NeighShrink square root thresholding method was proposed for solar radio spectrum image denoising [1]. Yihua Yan proposed a nonlinear calibration method to address the instrument saturation effect. The method adopted a channel normalization algorithm for image enhancement and finally used a wavelet approach to eliminate the effect of noise [2]. For the detection of the solar radio spectrum outline, Yan Zhang improved the level set method to extract the contours of the original image, overcoming the problem of missing detection associated with this method [3]. Chengming Tan developed a data analysis system compatible with SSW (Solar Soft Ware). This software can further expand the connection between data while mining the deep information of data [4]. White S.M. used data from the Green Bank Solar Radio Burst Spectrometer to describe and illustrate the main types of radio bursts [5]. In the study of sunspots, Preminger proposed a new method for the fast and automatic identification of solar features using the contrast ratio of images [6]. K. Iwai developed a new metric spectrum observation system for solar radio bursts. This system allows us to observe solar radio bursts in the frequency range of 150 to 500 MHz [7].

With the rapid development of deep learning, deep learning has achieved good results in image processing and computer vision [8,9]. Researchers have begun to use deep learning to study the solar radio spectrum. Vasili V. Lobzin proposed a method for the automatic identification and classification of type II and type III solar radio spectra. The method improved the accuracy of detection of solar radio burst types II and III via preprocessing, spectral intensity transformation, and spectral frequency transformation [10,11]. Junchen Guo proposed a hybrid network based on convolutional and GRU storage units to classify the solar radio spectrum [12]. Weidan Zhang proposed a model that combines a conditional generative adversarial network (CGAN) and a deep convolutional generative adversarial network (DCGAN) to classify the solar radio spectrum [13].

In previous studies, most researchers adopted convolutional operations as backbone networks to extract features of the solar radio spectrum. The methods had a large number of model parameters and consumed many system resources. In this paper, we propose a method combining a Swin transformer and transfer learning to classify solar radio spectra. Our method incorporates the advantages of convolutional neural networks, such as localization, translation invariance and hierarchy. Compared with previous research in classifying the solar radio spectrum, our method achieves better classification accuracy, while the number of model parameters is greatly decreased.

## 2. Solar Radio Spectrum and Preprocessing

*2.1. Dataset Introduction*

The dataset for this research experiment was obtained from the national astronomical observatory of the Chinese Academy of Sciences. The data were collected via a solar broadband spectrometer (SBRS). The sample amounts of the dataset are shown in Table 1, which are in references [12,13].

**Table 1.** Solar radio spectrum dataset.

| Spectrum Type | Burst | Nonburst | Calibration | Total |
| --- | --- | --- | --- | --- |
| Spectrum number | 579 | 3335 | 494 | 4408 |

The vertical axis of the solar radio spectrum represents the frequency of the spectrum, and the horizontal axis represents the time. Each pixel value represents the solar radio flux at a given time and frequency. If the spectrum images are displayed in grayscale, white represents a high flux of solar radio emission, and black represents a low flux of solar radio emission. The whole image represents the solar radio flux at a certain frequency during a certain time period. The burst, nonburst, and calibrated solar radio spectrum are shown in Figure 1a–c. Sometimes, the solar radio spectrum can be displayed with pseudocolor to generate a good visual effect of the solar radio spectrum.

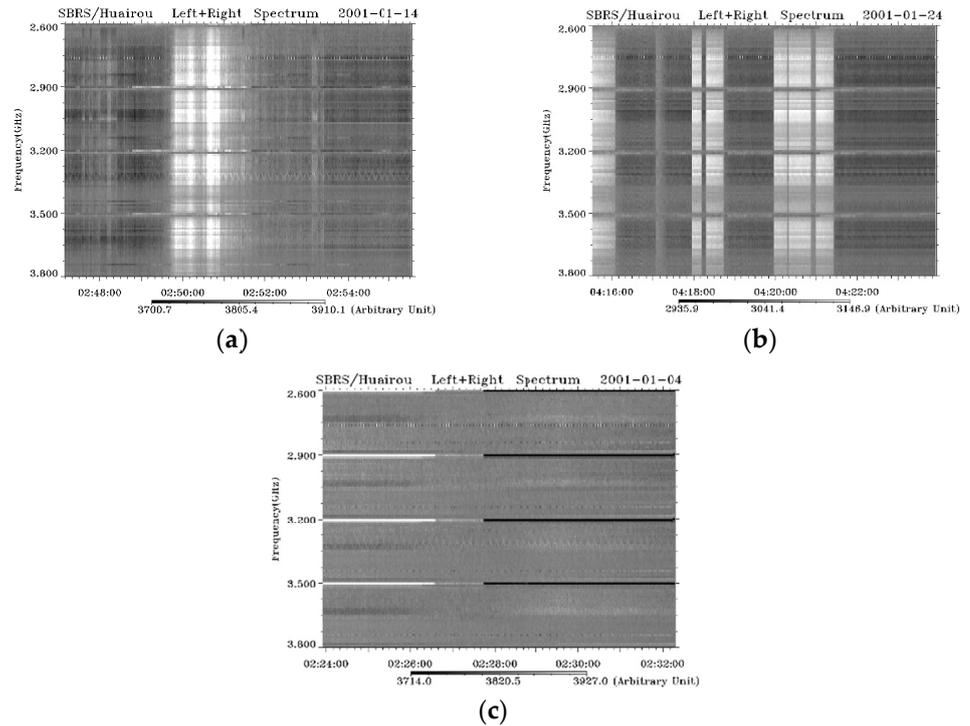

Figure 1. Solar radio spectrum: (**a**) "burst" type, (**b**) "calibrating" type, (**c**) "nonburst" type.

After the solar radio spectrometer receives solar radio emission, the signal is amplified and filtered and then saved in a certain file format. Horizontal stripes and other noise can appear in solar radio spectrum images due to internal interference, external interference, and quantization problems of the instrument. Therefore, it is necessary to preprocess the solar radio spectrum data.

*2.2. Normalization of Channels*

In a general sense, the nonlinear effect of the instrument is caused by the variation in the gain of each channel. There should be a deterministic continuous trend in the gain between channels. Therefore, we adopt the logic of the channel normalization method in the literature [12] to adjust the channel differences. That is, the pixel points on the horizontal stripes are subtracted from the original image $f(x,y)$, and the global average is added. The equation for the channel-normalized image $g(x,y)$ is

$$g(x,y) = f(x,y) - \frac{1}{n}\sum_{x=0}^{n} f(x,y) + \frac{1}{nm}\sum_{x=0}^{m}\sum_{y=0}^{n} f(x,y) \qquad (1)$$

where $f(x,y)$ is the pixel value of the original image at $(x,y)$, $m$ is the number of pixel points on the $x$-axis, and $n$ is the number of pixel points on the $y$-axis. The experimental results are shown in Figure 2a,b.

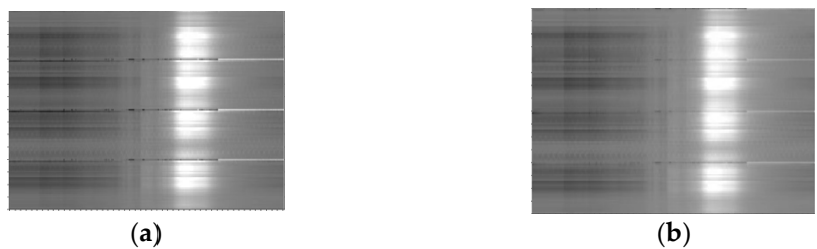

Figure 2. Normalized noise reduction results of channels: (**a**) before normalization of channels, (**b**) after normalization of channels.

*2.3. Pseudocolor Conversion and Dimensional Transformation of the Solar Radio Spectrum*

The Swin transformer network model requires the input format of the image to be a three-channel color map of 224 × 224 × 3. After channel normalization of the solar radio spectrum, pseudocolor conversion of the grayscale image and transformation of the image size are needed.

We define a grayscale pseudocolor mapping table to map the normalized grayscale values from the interval [0,1] to the corresponding pseudocolor. Figure 3a shows the pseudocolor color mapping table. The leftmost part of the color band corresponds to a grayscale of 0.0, the rightmost part corresponds to a grayscale of 1.0, and the middle corresponds to a grayscale of 0.5. With this grayscale pseudocolor mapping table, the channel-normalized Figure 2b can be converted to the pseudocolor image in Figure 3b.

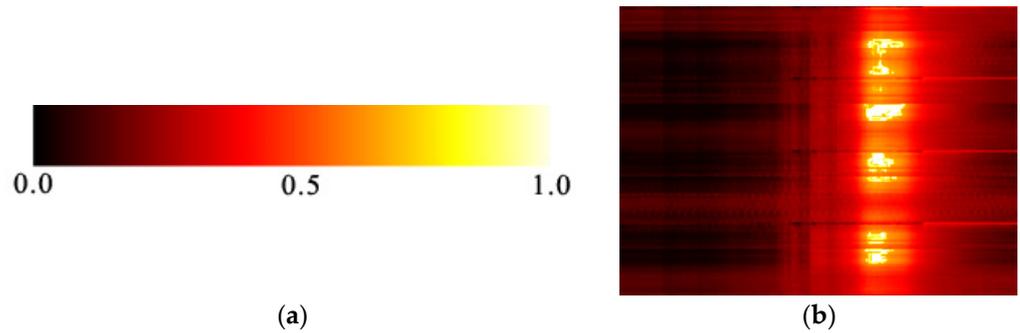

(**a**) (**b**)

**Figure 3.** Pseudocolor conversion results: (**a**) grayscale pseudocolor mapping table, (**b**) after pseudocolor.

After obtaining the pseudocolor image of the solar radio spectrum, we used a bilinear interpolation algorithm [16] to transform it to a size of (224 × 224 × 3). In the image size transformation, for any target pixel $g(x,y)$ after the transformation, the floating-point coordinates $(i+u, j+v)$ of the original image are obtained via the coordinate inversion transformation. Then, this pixel value $g(x,y)$ can be determined using the pixel values $f(i,j)$, $f(i+1,j)$, $f(i,j+1)$, and $f(i+1,j+1)$ of the four points around the original image $(i+u, j+v)$ together. The calculation formula is

$$g(x,y) = a+b+c+d \tag{2}$$

where

$$a = (1-u)(1-v)f(i,j) \tag{3}$$

$$b = (1-u)vf(i,j+1) \tag{4}$$

$$c = u(1-v)f(i+1,j) \tag{5}$$

$$d = uvf(i+1,j+1) \tag{6}$$

where $i$ and $j$ are nonnegative integers, and $u$ and $v$ are floating-point numbers in the interval [0,1). The RGB components of each pixel in the solar radio spectrum are processed separately using a bilinear interpolation algorithm to obtain the (224×224×3) solar radio pseudocolor spectrum.

## 3. Method

### 3.1. Transfer Learning

Solar activity depends heavily on the phase of the solar cycle. Solar radio bursts are low-probability events, and we found that the cumulative duration of solar radio bursts is less than 0.5% of the total observation duration after our statistics. Therefore, the number of solar radio burst spectrum samples collected in the last decade is limited. The number of solar radio bursts used in our experiments is 579, which is insufficient for training the network model.

To reduce the training cost and improve the training efficiency of the model, we use a transfer learning strategy. Transfer learning refers to the reuse of a pretrained model in another task. Usually, these pretrained models consume considerable time and computational resources in training, and transfer learning can transfer a pretrained model with acquired capabilities to a relevant problem. It is focused on finding similarities between existing knowledge and new knowledge, and transfer learning is achieved through transferring such similarities. Transfer learning is defined as follows:

Given a source domain $D_s$, a learning task $T_s$, a target domain $D_t$, and a learning task $T_t$, the purpose of transfer learning is to acquire knowledge in the source domain $D_s$ and the learning task $T_s$ to help enhance the learning of the prediction function $f_t(x)$ in the target domain, where $D_s \neq D_t$ or $T_s \neq T_t$. The model is shown in Figure 4 below:

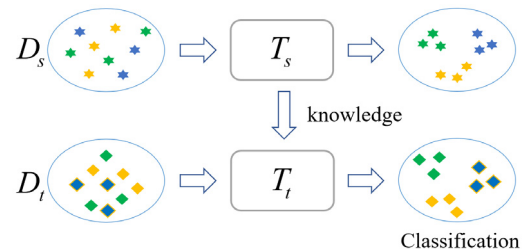

**Figure 4.** Transfer learning model.

At the beginning of training, the weights of the hidden layers of the pretrained Swin transformer model are first frozen. Then, the fully connected layer of the pretrained Swin transformer model is trained. After the above operation, the transfer learning process is completed.

### 3.2. Solar Radio Spectrum Classification Based on Swin Transformer

The Swin transformer, a deep learning network model proposed by Microsoft 2021, is a new network model with self-attention as the backbone network [14]. The Swin transformer constructs a hierarchical transformer [18] by using the hierarchical construction method commonly used in convolutional neural networks (CNNs). It also enacts the idea of self-attentiveness based on sliding windows to solve the problem of too much computational complexity caused by the transfer of the transformer to computer vision tasks. In addition, its design philosophy incorporates the essence of the residual network (ResNet) from the local to the whole, employing the transformer as a tool for progressively expanding the field of perception.

In contrast to a vision transformer (ViT) [18], a Swin transformer extracts the features of the image by continuously enlarging the attention window much like the convolution process of CNN. Its self-attention computation introduces the process of local aggregation, which is computed in terms of windows; the size of its window sliding step is equal to the window size to achieve window nonoverlap. When a traditional CNN performs convolutional operations on each window, each window takes on a value that represents the features of that window. In contrast to CNNs, Swin transformers calculate the self-attention of each window and obtains an updated window. Then, these windows are merged into a large window using a patch merge layer. Finally, in this large window, the value of self-attention is computed, and the obtained value represents the features of the entire window. Computing window-based self-attention in the window instead of global self-attention can greatly reduce the complexity of computation from $O(n^2)$ to $O(n)$. The complexity calculation formula is shown below:

$$\Omega(MSA) = 4hwC^2 + 2(hw)^2 C \tag{7}$$

$$\Omega(WMSA) = 4hwC^2 + 2(hw)^2 C \tag{8}$$

where $C$ is the number of channels in the image, $h$ and $w$ are the height and width of the image, respectively, and $M$ is the window size. Then, according to the self-attention formula, $MSA$ (global attention) $h \times w$ has $O(n^2)$ complexity, while WMSA (window-based attention) has $O(n)$ complexity when $M$ is fixed (window size, set to 7 by default). The larger the value of $h \times w$ is, the greater the time complexity of the global self-attention calculation and the more system resources will be consumed.

The architecture of the Swin transformer is very similar to that of the CNN. A Swin transformer contains four stages, each of which is a similar repetitive unit. First, the patch partition layer and the linear embedding layer divide $h \times w \times 3$ inputs into a collection of nonoverlapping patches, where the size of each patch is $4 \times 4$; then, the number of patches is $\frac{h}{4} \times \frac{w}{4}$. The classification process of the solar radio spectrum is shown in Figure 5.

In Figure 5, the solar radio spectrum is size $28 \times 28$ after image preprocessing, and the input image is transformed into patches that are composed of four adjacent pixel blocks by the patch partition layer and then fed into Stage 1. In Stage 1, the feature dimension of patches is changed to 96 by the linear embedding layer and then sent to the Swin transformer block. The input comes out of Stage 1 and enters the patch merging layer, where the input patches are merged with adjacent patches according to the rules of $2 \times 2$. The number of patches becomes $28 \times 28$, and the feature dimension becomes 192, which is equivalent to the image upsampling process. The dimension of the feature map decreases by half when passing through a stage, and the number of channels increases as the dimension decreases. Stages 2 to 4 follow the same principle, and the Swin transformer block in Stage 3 is cycled 6 times. The training process is similar to the convolution and pooling process in convolutional neural networks.

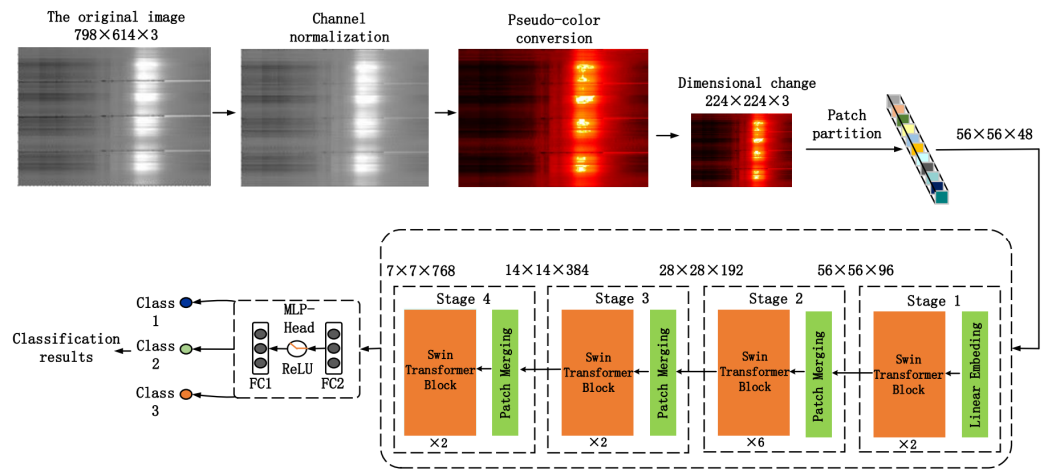

**Figure 5.** Solar radio spectrum classification process.

## 4. Experimentation and Discussion

We randomly selected a number of samples from the dataset in equal proportions for model pretraining, validation, and testing. The number of datasets for the training process is shown in Table 2 below:

**Table 2.** Experimental dataset division.

| Dataset | Burst | Nonburst | Calibration | Total |
|---|---|---|---|---|
| Training set | 200 | 1200 | 200 | 1600 |
| Validation set | 179 | 935 | 94 | 1208 |
| Test set | 200 | 1200 | 200 | 1600 |

### 4.1. Experimental Evaluation Metrics

We evaluated the model using true positive rate (TPR) and false positive rate (FPR) as experimental metrics, which can reflect the effectiveness of experimental classification. Among them, TPR is defined as the proportion of correctly detected positive samples to all positive samples; FPR is defined as the proportion of incorrectly detected false-positive samples to all negative samples. TPR and FPR are calculated as follows:

$$TPR = \frac{TP}{(TP+FN)} \tag{9}$$

$$FPR = \frac{FP}{(FP+TN)} \tag{10}$$

The predicted and the labeled values of the images are used as parameters of the confusion matrix algorithm to obtain TP (true-positive), TN (true-negative), FP (false-positive), and FN (false-negative). In this experiment, we expect that the higher the TPR, the better, and the lower the FPR, the better.

### 4.2. Experimental Results and Analysis

This research used the TensorFlow 2.7 deep learning framework on Python 3.8 to complete the experiment, with 32 GB of memory, a 2080 Ti graphics card, a 14,000 MHz memory frequency, an 11 GB memory capacity, a 352-bit memory bit width, a 0.001 learning rate, a cross-entropy loss function, and the activation function softmax. In this research, we froze the weights of the hidden layer of the pretrained model Swin transformer and then trained the fully connected layer of the Swin transformer on the target dataset. Finally, we saved the best parameters of the model for the following data testing. To ensure that the proposed model is efficient in the training of the entire solar radio spectrum and is not affected by the target dataset, we randomly upset all the training datasets at the beginning of each training session. The obtained experimental results are shown in Figure 6a,b and Table 3 below:

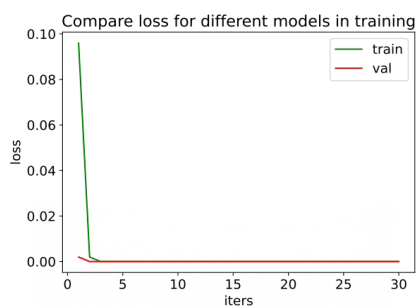
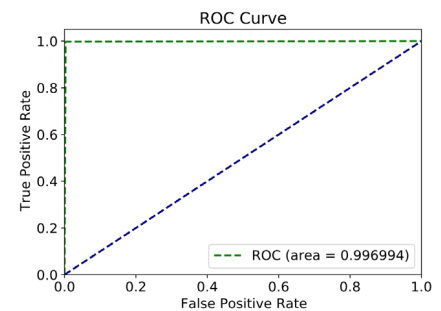

(**a**) (**b**)

**Figure 6.** Model training results: (**a**) visualization result of the loss value, (**b**) ROC curve.

**Table 3.** Experimental results of Swin transformer.

| Model | Swin Transformer | | Swin Transformer+ Transfer Learning | |
|---|---|---|---|---|
| | TPR (%) | FPR (%) | TPR (%) | FPR (%) |
| Burst | 98.9 | 0 | 100 | 0 |
| Nonburst | 100 | 0.5 | 100 | 0 |
| Calibration | 99.5 | 0.1 | 100 | 0 |

As seen from Figure 6a, the proposed method combining a Swin transformer and transfer learning has a loss value of 0 at the beginning of the third training round, and it remains at 0 in each subsequent training round, which fully demonstrates the superiority of our proposed algorithm.

As shown in Table 3, Swin transformer transfer learning can improve the classification efficiency of the model with the same proportion of dataset segmentation. Compared with experiments without transfer learning, the TPR of burst improves by 1.1%; the TPR of calibration improves by 0.5%; the FPR of nonburst decreases by 0.5%; and the FPR of calibration decreases by 0.1%. We find that using a transfer learning strategy can improve the classification effect of the model well.

In terms of model training parameters, this design was compared with the previous large model VGG16. We find that the training parameters of the Swin transformer decreased by approximately 80% compared to the VGG16 convolutional neural network without reducing the TPR and FPR. The results are shown in Table 4.

**Table 4.** Swin transformer vs. VGG16.

| Model | Swin Transformer+ Transfer Learning | | VGG16+ Transfer Learning | |
|---|---|---|---|---|
| | TPR (%) | FPR (%) | TPR (%) | FPR (%) |
| Burst | 100 | 0 | 96.8 | 1.4 |
| Nonburst | 100 | 0 | 97.1 | 1.3 |
| Calibration | 100 | 0 | 99.6 | 1.8 |
| Parameters | 27, 550, 473 | | 139, 357, 544 | |

To demonstrate the advantages of our proposed method, we compare the model parameters of the vision transformer, which extracts image features similarly to the Swin transformer. Without reducing the TPR and FPR, the number of parameters of the Swin transformer decreases by approximately 60% compared to the vision transformer. The results are shown in Table 5.

**Table 5.** Swin transformer vs. vision transformer.

| Method | Swin Transformer+ Transfer Learning | | Vision Transformer | |
|---|---|---|---|---|
| | TPR (%) | FPR (%) | TPR (%) | FPR (%) |
| Burst | 100 | 0 | 99.5 | 0 |
| Nonburst | 100 | 0 | 100 | 0 |
| Calibration | 100 | 0 | 100 | 0.1 |
| Parameters | 27, 550, 473 | | 85, 800, 963 | |

To demonstrate the advantage of our proposed method when training the model, we also compared the training time of the VGG16 model, the vision transformer model, and the Swin transformer model. The results are shown in Figure 7 below.

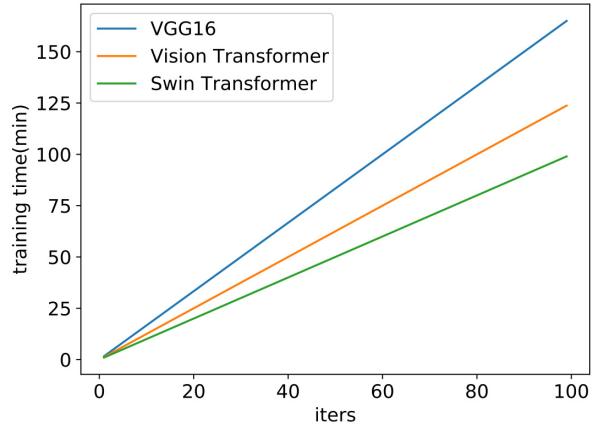

**Figure 7.** Comparison of training time for each model.

Experimental comparison demonstrates that the proposed method combining the Swin transformer and transfer learning obtains excellent results for the classification of solar radio spectrum. Not only does the model training converge more stably, but the training time is also shorter. Finally, we compared the experimental results of references [12,13,16,18] and the current method, as shown in Table 6.

**Table 6.** Comparison results of all experiments.

| Model | | burst | nonburst | calibration |
|---|---|---|---|---|
| Swin transformer | TPR (%) | 100 | 100 | 100 |
| | FPR (%) | 0 | 0 | 0 |
| Vision transformer | TPR (%) | 99.5 | 100 | 100 |
| | FPR (%) | 0 | 0 | 0.1 |
| CGRU | TPR (%) | 96.8 | 99.5 | 99.9 |
| | FPR (%) | 0 | 1.5 | 0.3 |
| VGG16 | TPR (%) | 96.8 | 97.1 | 99.6 |
| | FPR (%) | 1.4 | 1.3 | 1.8 |
| CNN | TPR (%) | 84.6 | 90 | 99 |
| | FPR (%) | 9.7 | 8.7 | 0.7 |
| Multimodel | TPR (%) | 70.9 | 80.9 | 96.8 |
| | FPR (%) | 15.6 | 13.9 | 3.2 |
| DBN | TPR (%) | 67.4 | 86.4 | 95.7 |
| | FPR (%) | 3.2 | 14.1 | 0.4 |
| PCA+SVM | TPR (%) | 52.7 | 0.1 | 68.3 |
| | FPR (%) | 2.6 | 16.6 | 72.2 |

According to Table 6, all of our experimental results show a significant improvement over those of previous researchers. This gain can be attributed to the advantages of the Swin transformer in image processing, which draws on the features of Resnet for extracting image features and employs the transformer as a tool to gradually extend the perceptual domain from local to global.

**5. Conclusions**

In this paper, we propose a solar radio spectrum classification method combining a Swin transformer and transfer learning. Experiments show that the self-attentive mechanism can extract the global features of images well, which gives the model a strong generalization ability and greatly improves model classification. This paper can provide reference for other astronomical image classification.

**Author Contributions:** J.C. proposed the network architecture design and the framework of a Swin transformer. L.Y. and S.L. collected and preprocessed the datasets. J.C. performed the experiments. J.C., L.Y., and S.L. analyzed and discussed the experimental data. J.C. and G.Y. wrote the article. H.Z., and C.T. revised the article and provided valuable advice for the experiments. All authors have read and agreed to the published version of the manuscript.

**Funding:** This research was funded by the Natural Science Foundation of China (Grant No. 12263008, 11941003), the MOST (Grant No. 2021YFA1600500), the Application and Foundation Project of Yunnan Province (Grant No. 202001BB050032), the Yunnan Provincial Department of Science and Technology–Yunnan University Joint Special Project for Double-Class Construction (Grant No. 202201BF070001-005), the Key R&D Projects of Yunnan Province (Grant No. 202202AD080004), the Open Project of CAS Key Laboratory of Solar Activity, National Astronomical Observatories (Grant No. KLSA202115), and the 13th Postgraduate Innovation Project of Yunnan University (Grant No. 2021Y269).

**Institutional Review Board Statement:**

**Informed Consent Statement:**

**Data Availability Statement:** The data are available at GitHub: https://github.com/filterbank/spectrumcls.

**Acknowledgments:** We would like to thank the anonymous reviewers and the editor-in-chief for their comments to improve the article. We give thanks also to the data sharer. We thank all the people involved in the study.

**Conflicts of Interest:** The authors declare no conflict of interest.

**References**

1. Ruizhen Zhao, Zhanyi Hu. Wavelet NeighShrink Method for Grid Texture Removal in Image of Solar Radio Bursts. Spectroscopy and Spectral Analysis, vol. 27, no. 1, p. 4, 2007. (in Chinese)
2. Yihua Yan, Chengming Tan, Long Xu, Huirong Ji, Qijun Fu, Guoxiang Song. Nonlinear Relative Calibration Methods and Data Processing for Solar Radio Bursts. Science of China: Mathematics, 2001. (in Chinese)
3. Yan Zhang, Yihua Yan, Chengming Tan, et al. Automatic Contour Detection and Information Extraction of Solar Radio Spectrogram. Modern Electronic Technology, vol. 34, no. 2, p. 4, 2011. (in Chinese)
4. Chengming Tan, Yihua Yan, Baolin Tan, et al. Design of A Data Processing System for Solar Radio Spectral Observations. Astronomy Research and Technology, 2011. (in Chinese)
5. White S M. Solar Radio Bursts and Space Weather. Asian Journal of Physics, 2007, 16: 189-207. (in Chinese)
6. Preminger D.G, Walton S.R, Chapman G.A. Solar Feature Identification using Contrasts and Contiguity. Solar Physics, 202（1）, pp53-62, 2001.
7. Iwai K, Tsuchiya F, Morioka A, et al. IPRT/AMATERAS: A new metric spectrum observation system for solar radio bursts. Solar Physics, 2012, 277(2): 447-457.
8. Liu H.; Yuan, G.; Yang, L.; Liu, K.; Zhou, H.. An Appearance Defect Detection Method for Cigarettes Based on C-CenterNet. Electronics, Vol. 11(14), 2182, 2022.
9. Yang L, Yuan G, Zhou H, Liu H, Chen J, Wu H.. RS-YOLOX: A High-Precision Detector for Object Detection in Satellite Remote Sensing Images. Applied Sciences, Vol.12(17):8707, 2022.
10. Lobzin V. V. , Cairns I H , Robinson P A , et al. Automatic recognition of type III solar radio bursts: Automated Radio Burst Identification System method and first observations. Space Weather-the International Journal of Research & Applications, 7(4), 2009.
11. Vasili V. Lobzin, Iver H. Cairns, Peter A. Robinson, etc. Automatic 19 recognition of coronal type II radio bursts: the automated radio burst identification system method and first observations. The Astrophysical Journal Letters, 710(1), pp58–62, 2010.
12. 12.    Juncheng Guo, Fabao Yan, Gang Wan, et al.. A deep learning method for the recognition of solar radio burst spectrum. PeerJ Computer Science 8:e855 https://doi.org/10.7717/peerj-cs.855
13. Weidan Zhang, Fabao Yan, Fuyun Han, et al. Auto recognition of solar radio bursts using the C-DCGAN method. Front. Phys. 9:646556. doi: 10.3389/fphy.2021.646556

14. 14. Ze Liu, Yutong Lin, Yue Cao, et al. Swin Transformer: Hierarchical vision transformer using shifted windows//Proceedings of the IEEE/CVF International Conference on Computer Vision. 2021: 10012-10022.
15. Sen Wang, Kejian Yang. Research and implementation of image scaling algorithm based on bilinear interpolation. Automation Technology and Applications, 2008, 27(7): 44-45. (in Chinese)
16. Sisi Chen. Research on Classification Algorithm of Solar Radio Spectrum Based on Convolutional Neural Network. Shenzhen University, 2018. (in Chinese)
17. Trockman A, Kolter J Z. Patches Are All You Need? arXiv preprint arXiv:2201.09792, 2022.
18. Min Chen, Guowu Yuan, Hao Zhou, et al. Classification of Solar Radio Spectrum Based on VGG16 Transfer Learning//Chinese Conference on Image and Graphics Technologies. Springer, Singapore, 2021: 35-48.